\documentclass[11pt]{article}
\newfont{\bb}{msbm10}

\usepackage{amsfonts}
\usepackage{amsmath,amssymb,amsthm}
\usepackage{booktabs}
\usepackage{amsbsy}
\usepackage{comment}
\usepackage{graphicx,color,overpic}
\usepackage[numbers,sort&compress,square]{natbib}
\usepackage{soul}

\def\Div{{\rm div}\,}
\newcommand{\bu} {\mathbf{u}}

\newcommand{\bn} {\mathbf{n}}

\newcommand{\bw} {\mathbf{w}}

\newcommand{\bx} {\mathbf{x}}
\newcommand{\ba} {\mathbf{a}}
\newcommand{\be} {\mathbf{e}}
\newcommand{\blf} {\mathbf{f}}

\newcommand{\QQ} {\mathbb{Q}}
\newcommand{\VV} {\mathbb{V}}

\newcommand{\bI} {\mathbf{I}}

\newcommand{\bD} {\mathbf{D}}
\newcommand{\bG} {\mathbf{G}}
\newcommand{\dO} {{\partial\Omega}}

\newcommand{\bB} {{\bf B}}

\newcommand{\bsigma}{\mbox{\boldmath$\sigma$\unboldmath}}
\newcommand{\bpsi}{\mbox{\boldmath$\psi$\unboldmath}}

\newcommand{\bxi}{\mbox{\boldmath$\xi$\unboldmath}}
\newcommand{\balpha}{\mbox{\boldmath$\alpha$\unboldmath}}

\begin{document}

\thispagestyle{empty}
\date{}

\title{A stable method for 4D CT-based CFD simulation in the right ventricle of a TGA patient}

\author{
Alexander Danilov \thanks{Marchuk Institute of Numerical Mathematics RAS and Moscow Institute of Physics and Technology; {\tt a.a.danilov@gmail.com}}
\and
Yushui Han
\thanks{DeBakey Heart\&Vascular Center, Houston Methodist Hospital; {\tt yhan@houstonmethodist.org}}
\and
Huie Lin
\thanks{DeBakey Heart\&Vascular Center, Houston Methodist Hospital; {\tt clin@houstonmethodist.org}}
\and
Alexander Lozovskiy\thanks{Marchuk Institute of Numerical Mathematics RAS; {\tt saiya-jin@yandex.ru}}
\and
Maxim A. Olshanskii\thanks{Department of Mathematics, University of Houston; {\tt molshan@math.uh.edu}}
\and
Victoria Yu. Salamatova\thanks{Sechenov University and Moscow Institute of Physics and Technology; {\tt salamatova@gmail.com}}
\and
Yuri V. Vassilevski\thanks{Marchuk Institute of Numerical Mathematics RAS, Sechenov University,  and Moscow Institute of Physics and Technology; {\tt yuri.vassilevski@gmail.com}}
}

\maketitle

\markboth{}{CFD for right ventricle with TGA}

\begin{abstract} The paper discusses a stabilization of a finite element method for the equations of fluid motion in a time-dependent domain. After experimental convergence analysis, the method is applied to simulate a blood flow in the right ventricle of a post-surgery patient with the transposition of the great arteries disorder. The flow domain is reconstructed from a sequence of  4D CT images. The corresponding segmentation and triangulation algorithms are also addressed in brief.
\end{abstract}




\section{Introduction}\label{intro}
Last decades evidenced a remarkable progress in the development of mathematical and computational models for physiological flows and
their interaction with surrounding tissues and organs. In particular, computer simulations of the human  cardiovascular system and its parts
have been a focus of intense research. We refer to the recent monographs~\cite{tu2015computational,quarteroni2019mathematical,vassilevski2020personalized} for the overview of the field.
Based on these advances, the CFD visualization and quantification of blood flow in the heart
and large vesicles has a potential to become a clinical standard and to complement Doppler sonography as a decision supporting tool
for practicing cardiologists~\cite{Munoz2013}.
Nevertheless, reliable and predictive patient-specific  simulations of  flow in the heart chambers
remains a challenge, especially under pathological or post-surgery conditions.
For example, for the CFD reconstruction of the flow in a heart ventricle, the challenge consists of quality image acquisition, image segmentation,
recovery of tissue motions, a suitable volume tessellation (meshing), a  discretization of the system of governing partial differential equations,
fast solution methods and postprocessing of the computed solution.
While the present paper briefly addresses several of these stages of personalized CFD simulations,
the focus here is on building a stable discrete approximation of the Navier--Stokes equations for a transitional flow in an evolving domain.
A concrete practical problem we are interested in here is the simulation of   blood flow in the right ventricle reconstructed from 4D CT images of the heart of a patient with transposition of the great arteries~(TGA).

The rest of the paper is organized in five sections. In section~\ref{sec:model} we introduce a system of the Navier--Stokes equations
governing the motion of incompressible viscous fluid (blood) in a time-dependent domain (the heart chamber) with appropriate boundary conditions.
Section~\ref{sec_FE} discusses a discretization method with particular attention to spatial stabilization.  Section~\ref{sec_seg} adds details of the medical image processing and meshing techniques.
Section~\ref{s_num_examples} splits into two parts. In the first part, we show the results of convergence tests for the numerical method,
which confirm the expected accuracy rates. The second part presents the CFD visualization and analysis of flow in the right ventricle.

\section{Mathematical model}\label{sec:model}
We represent the heart chamber as a 3D time-dependent domain $\Omega(t)\subset\mathbb{R}^3$
occupied by a viscous fluid (blood) for the simulation time $t\in[0,T]$.
We assume that the deformation of $\Omega(t)$ is smooth enough in the sense that there exists a two times continuously differentiable
one-to-one mapping $\bxi$ from the reference  domain $\Omega_0=\Omega(0)$ to the physical domain, i.e.,
$\bxi~:~\Omega_0\to\Omega(t)\quad\text{for}~t\in[0,T]$
and
$\bxi\in C^{2}([0,T]\times\overline{\Omega}_0)$.

For the blood flow in the heart,  it is reasonable to assume that the fluid is Newtonian and incompressible~\cite{vassilevski2020personalized}.
The dynamics of incompressible Newtonian fluid is governed by the system of  Navier--Stokes equations,
\begin{equation}\label{NSE0}
\left\{
\begin{aligned}
\mathrm{\frac{d}{dt}}\bu- 2\nu\Div(\bD(\bu))+\nabla p  &=\blf\\
 \Div\bu&= 0
 \end{aligned}\right. \quad\text{in}~~\Omega(t),~~\text{for}~t\in(0,T),
 \end{equation}
written for the unknown fluid velocity vector field $\bu(\bx,t)$ and the unknown pressure function $p(\bx,t)$.
In equation \eqref{NSE0},  $\nu$ is the kinematic viscosity coefficient,
$\bD(\bu)$ stands for the rate of deformation tensor, and
$\mathrm{\frac{d}{dt}}$ denotes the material (Lagrangian) derivative, i.e. the derivative along material trajectories of particles.
For numerical purposes, it is convenient to expand the material derivative in Eulerian terms and
to re-write the fluid system in the so-called arbitrary Lagrangian--Eulerian (ALE) form,
\begin{equation}\label{NSE}
\left\{
\begin{aligned}
\left.\frac{\partial \bu}{\partial t}\right|_{\Omega_0}+((\bu-\bw)\cdot\nabla)\bu- 2\nu\Div(\bD(\bu))+\nabla p  &=\blf\\
 \Div\bu&= 0
 \end{aligned}\right. \quad\text{in}~~\Omega(t),~~\text{for}~t\in(0,T).
 \end{equation}
 Here $\displaystyle\bw:=\frac{\partial \bxi(t,\bxi^{-1}(\bx))}{\partial t}$ is the ALE velocity at  $\bx\in\Omega(t)$ and
 $\displaystyle\left.\frac{\partial \bu}{\partial t}\right|_{\Omega_0}:=\frac{\partial \hat\bu(t,\bxi^{-1}(\bx))}{\partial t}$,
 with  $\hat\bu:=\bu\circ\bxi$, is the time derivative of velocity in the reference frame. Further we shall write $\bu_t$ for the sake of notation simplicity.

 The system~\eqref{NSE} should be supplemented with boundary conditions for velocity or stress.
 These boundary conditions drive the flow and so they are important part of the model.
 On the walls of the chamber, denoted by $\dO^{\rm ns}(t)$, we set no-penetration and no-slip boundary condition for fluid, which for the moving domain take the following form:
 \begin{equation}\label{cont_int}
\bu=\bxi_t\circ\bxi^{-1}\quad\mbox{on}~\dO^{\rm ns}(t).
\end{equation}
 The part of the boundary corresponding to the tricuspid valve is designated as the inflow part, $\dO^{\rm in}(t)$, with the following boundary conditions
\begin{equation}\label{bc2}
\bu =\bxi_t\circ\bxi^{-1}\quad\mbox{on}~\dO^{\rm in}(t),~\mbox{for systolic phase}~~ \mbox{and}~~ \bsigma\bn=0\quad\mbox{on}~\dO^{\rm in}(t),~\mbox{for diastolic phase},
\end{equation}
where $\bsigma=-2\nu\bD(\bu)+p\bI$ is the Cauchy stress tensor, $\bn$ is the unit normal vector on the boundary of $\Omega(t)$.
These conditions imply that the blood freely flows in the ventricle through the tricuspid valve driven by the compartment expansion.
The part of the boundary with the pulmonary valve represents the outflow part, $\dO^{\rm out}(t)$, with the boundary condition
\begin{equation}\label{bc3}
\bsigma\bn =0\quad\mbox{on}~\dO^{\rm out}(t),~\mbox{for systolic phase}~~ \mbox{and}~~ \bu=\bxi_t\circ\bxi^{-1}\quad\mbox{on}~\dO^{\rm out}(t),~\mbox{for diastolic phase}.
\end{equation}
Finally, we need to define an initial state of the system, which in the absence of other data we assume to be the fluid at rest, $\bu=0$ in $\Omega(0)$.
One may need to simulate several cardiac cycles to obtain $\bu$ and $p$ non-sensitive to the error induced by  such non-physiological  initial state.

To build a discrete model, we first need an integral formulation of the  system. We multiply the first equation in \eqref{NSE} by a smooth vector function
$\bpsi:\Omega(t)\to \mathbb{R}^3$ such that $\bpsi=\mathbf{0}$ on $\dO^{\rm ns}(t)\cup\dO^{\rm in}(t)$ during systolic phase  and
$\bpsi=\mathbf{0}$ on $\dO^{\rm ns}(t)\cup\dO^{\rm out}(t)$
during diastolic phase. Integrating the resulting identity over $\Omega(t)$ and by parts we get
 \begin{multline}\label{weak}
\int_{\Omega(t)} \{\bu_t+((\bu-\bw)\cdot\nabla)\bu\}\cdot\bpsi\,dx +2\nu\int_{\Omega(t)}\bD(\bu):\bD(\bpsi)\,dx \\ -\int_{\Omega(t)}p\Div\bpsi\,dx +\int_{\Omega(t)}q\Div\bu\,dx=\int_{\Omega(t)}\blf\cdot\bpsi\,dx,
\end{multline}
where we used boundary conditions and added the second equation in \eqref{NSE} tested by a smooth function $q:\Omega(t)\to \mathbb{R}$.
For a regular solution to  \eqref{NSE}--\eqref{bc3}, i.e. for smooth velocity field $\bu$ and pressure $p$,
the identity \eqref{weak} holds for any $t\in(0,T)$ and  any smooth test functions $q$ and $\bpsi$ as specified above.

\section{Discrete models}\label{sec_FE}

We start with building a triangulation of the reference domain $\Omega_0$: Let
$\mathcal{T}_h$ be  a collection of  tetrahedra such that $\overline{\Omega}_0=\bigcup_{T\in\mathcal{T}_h}\overline{T}$
and any two tetrahedra from $\mathcal{T}_h$ intersect by either an entire face, an entire edge, a vertex, or the empty set.
We also assume that the mesh is regular in the sense that the minimal angle condition holds for all tetrahedra from $\mathcal{T}_h$, cf., e.g.,~\cite{ciarlet2002finite}.
A triangulation satisfying all above conditions is called admissible. In turn, triangulations $\mathcal{T}_h(t)$ of $\Omega(t)$ are built using $\bxi(t)$ to map tetrahedra vertices from $\Omega_0$ to $\Omega(t)$.
The mapping  $\bxi$ should be such  that the resulting sequence of meshes has the same connectivity and delivers admissible triangulations of  $\Omega(t)$.
Consider conforming FE spaces $\VV_h(t)\subset H^1(\Omega(t))^d$  and $\QQ_h(t)\subset L^2(\Omega(t))$,
spaces of continuous  piecewise polynomial functions on $\mathcal{T}_h(t)$;
$\VV_h^0$ is a subspace of $\VV_h$ of functions vanishing on the same part of the boundary as the test function from the integral formulation~\eqref{weak}.
In this paper, we choose the Taylor--Hood (P2/P1) pair of finite element spaces for the  velocity--pressure pair~\cite{hood1974navier}:
\begin{equation}\label{defVQ}
\begin{aligned}
\VV_h(t)&=\{ \bu_h \in C(\Omega(t))^3\,:\, \bu_h|_T \in \left[P_{2}(T)\right]^3, \forall~ T \in \mathcal{T}_h(t)\}, \\
\QQ_h(t)&=\{ q_h \in C(\Omega(t))\,:\, q_h|_T \in P_{1}(T), \forall~ T \in \mathcal{T}_h(t)\},
\end{aligned}
\end{equation}
which are known to satisfy the necessary inf-sup stability condition~\cite{bercovier1979error}.
The mapping $\bxi$ is also approximated with a piecewise polynomial mapping $\bxi_h\in\VV_h(0)$, which is constructed by the interpolation of $\bxi$ using its nodal values.

We now consider discretization in time. Assume a constant time step $\Delta t=\frac{T}{N}$, where $N$ is the total number of steps.
We use the notations $t_k=k\Delta t$, $\bu^k:=\bu(t_k, \bx)$, and similar for $p$, $\bxi$, and  $\Omega_k=\Omega(t_k)$.
For a sequence of functions $f^i$, $i=0,\dots,k$, all defined in the reference domain,  $\left[ f\right]^{k}_t:=\frac{f^k-f^{k-1}}{\Delta t}$
denotes the backward finite difference at $t_k$. Let $\bu_h^0$ be the Lagrange interpolant of  the initial velocity field.
The fully discrete problem builds on the integral formulation  \eqref{weak} and reads:
For $k=1,2,\dots$, find $\bu^{k}_h\in \VV_h(t_k)$, $p^{k}_h\in \QQ_h(t_k)$ satisfying Dirichlet boundary conditions in \eqref{cont_int},\eqref{bc2},\eqref{bc3},
and the integral equality
\begin{multline}\label{FE1}
\int_{\Omega_k} \{\left[\widehat{\bu}_h\right]_t^{k}\circ(\bxi^k_h)^{-1}
+((\widetilde\bu_h^{k-1}-\bw_h^k)\cdot\nabla)\bu_h^k\}\cdot\bpsi_h\,dx +2\nu\int_{\Omega_k}\bD(\bu^k_h):\bD(\bpsi_h)\,dx \\ -\int_{\Omega_k}p^k_h\Div\bpsi_h\,dx +\int_{\Omega_k}q_h\Div\bu^k_h\,dx=\int_{\Omega_k}\blf\cdot\bpsi_h\,dx,
\end{multline}
for all $\bpsi_h\in\VV_h^0(t_k),$ $q_h\in\QQ_h(t_k)$.
Here the advection velocity is computed with the help of the mapping to the reference domain,
$\bw_h^k(\bx)=\left[\bxi_h\right]_t^{k}((\bxi^k_h)^{-1}(\bx))$, as well as
$\widetilde\bu_h^{k-1}(\bx):=\bu_h^{k-1}(\bxi^{k-1}_h(\bxi^k_h)^{-1}(\bx))$, for $\bx\in \Omega^k$, and
$\widehat\bu_h^k:=\bu_h^k\circ\bxi^k_h$.

Note that the inertia terms are linearized so that a \emph{linear} algebraic system should be solved on each time step.

Blood flow of the  heart ventricles is characterized by transitional or even turbulent regimes,
see, e.g., \cite{querzoli2010effect,falahatpisheh2012high,chnafa2014image}.
Therefore, sufficiently fine  resolution of spatial and time scales in $\bu$ is required for the direct numerical simulation of  blood flows in the  ventricles.
Adopting such resolution would lead to computations prohibitively expensive for patient specific modelling.
An alternative to employing very fine triangulation and time stepping is the use of  a  Large Eddy Simulation (LES) model of turbulence~\cite{chnafa2014image}
or another subgrid method designed to model the effect of unresolved scales and to dissipate excessive energy.
Such models aim  to deliver  stable numerical approximations for the mean flow statistics. This is the approach taken in the present paper.
More specifically, we consider a combination of the classical Smagorinski LES model~\cite{smagorinsky1963general} and
the streamline upwinded Petrov--Galerkin (SUPG) method~\cite{brooks1982streamline}.

SUPG method introduces subgrid modelling on the level of discretization~\cite{hughes2018multiscale}.
The variant we use consists in adding to \eqref{FE1} the element-wise residual term of the form
\begin{equation}\label{SUPG}
\sum_{T\in\mathcal{T}_h(t_k)} \tau_T \int_T
 (\ba^k_h\cdot\nabla)\bpsi_h\left\{\left[\widehat{\bu}_h\right]_t^{k}+(\ba^k_h\cdot\nabla)\bu_h^k - 2\nu\Div(\bD(\bu_h^k))+\nabla p_h^k-\blf\right\}\,dx,
\end{equation}
with $\ba^k_h:=\widetilde\bu_h^{k-1}-\bw_h^k$,

The parameters $\tau_T$ are defined elementwise as follows
\[
\tau_T = \frac{1}{\sqrt{d_1 + d_2 + d_3}},
\]
with
\[
d_1 = 60\nu^2\mbox{tr}(\bG\bG^T),\quad
d_2 = \frac{4}{|\Delta t|^2},\quad
d_3 = (\ba_h^{k})^T\bG\ba_h^{k}.
\]
Matrix
$\bG\in\mathbb{R}^{3\times3}$, is the element metric tensor defined as $\bG = \bB^T\bB$, where $\bB=\frac{\partial \balpha}{\partial\bx}\in\mathbb{R}^{4\times3}$, and $\balpha$ is the mapping from the bulk to the barycentric coordinates of $T$.

The SUPG method is sufficiently accurate, since it is a residual type method.
However, we found SUPG formulation alone not always sufficiently stable for flow regimes typical to the ventricles on acceptable computational grids.
Therefore, we suggest to increase the robustness of the SUPG method by combining it with the   Smagorinski model.
In the  Smagorinski approach the effect of unresolved scales is modeled through introducing additional (turbulent) viscosity in the equation.
The turbulent viscosity depends on the local strain rate and adds to the physical viscosity.
The  Smagorinski model is relatively simple, but  known to be excessively dissipative, especially near walls~\cite{zang1993dynamic}.
We incorporate it in SUPG formulation by replacing $\nu$ in \eqref{FE1} with
\begin{equation}\label{turb_visc}
\nu_T = \nu + M\cdot(0.2 h_T)^2 \sqrt{2\bD(\widetilde\bu^{k-1}_h) : \bD(\widetilde\bu^{k-1}_h)},
\end{equation}
where  $h_T = \textrm{diam} (T)$, for any tetrahedral cell $T\in \mathcal{T}_h$.
Factor $M$  regulates how much turbulent viscosity we add to the method. Our goal is to set $M$ to a minimal possible value, which
makes SUPG stabilization robust for flow regimes typical for the ventricles.
Modification of viscosity increases the finite element error, yet the method is consistent in the sense that the added viscosity vanishes as $h_T\to 0$.

\section{Image segmentation and mesh generation}\label{sec_seg}
In this work, we apply the finite element CFD simulation for personalized modelling of the flow in the right ventricle of a patient
with transposition of the great arteries~(TGA), a rare congenital defect. The motion of the heart is reconstructed from 4D CT images. The incoming data set is a series of 10 contrast enhanced CT images with $512 \times 512 \times 304$ voxels and $0.355 \times 0.355 \times 0.5$ mm resolution. We crop  and resample each CT image to $162 \times 112 \times 136$ voxels with 1 mm isotropic resolution. The right ventricle is manually segmented for each of the ten CT images using a semi-automatic level-set method from ITK-SNAP package~\cite{py06nimg}.

Since the temporal resolution of ten images per cardiac cycle is not enough for smooth mesh transition, we also perform  temporal resampling.
For each frame we convert the binary mask of the segmented right ventricle to a signed distance function,
which is negative inside the ventricle, positive outside, and zero at the ventricle boundary. These ten scalar frames are resampled to 90 frames per cycle using a cubic  interpolation in time with the periodic conditions at the end points of the time interval. For each new frame, the boundary of the right ventricle is recovered  as a zero isosurface of this interpolant. Thus, the set of 90 frames represent a periodic motion. We next apply a cyclic shift of the frame indices to set the starting frame to be the one with the maximum volume of the ventricle. Then the frame 44 shows  the minimum volume of the ventricle. We assume that frames from 0 to 44 represent systole of the cardiac cycle, and frames from 45 to 89 represent diastole.

We manually select two static cutoff planes to represent the position of valves in the right ventricle.
We note that actual valves are moving during the cardiac cycle and do not stay in the same planes.
Therefore, we estimate the average position of the valves based on the available sparse temporal resolution.

The mesh generation process is similar to the one proposed in our previous work~\cite{Danilov_2017} and we refer to that paper for any omitted details.  The algorithm requires a reference domain, which is defined implicitly as an enclosed volume of the averaged  distance function over all 90 frames. This volume is also bounded by static valve planes.
A reference quasi-uniform unstructured tetrahedral mesh is first constructed with the help of  Delaunay triangulation algorithm from CGAL Mesh library~\cite{Rineau_2007}. We next improve the reference mesh quality using aniMBA library from the Ani3D package~\cite{Ani3D} and also enforce each tetrahedron to have at least one internal node.

\begin{figure}
\centering
    (a)\includegraphics[width=1.44in]{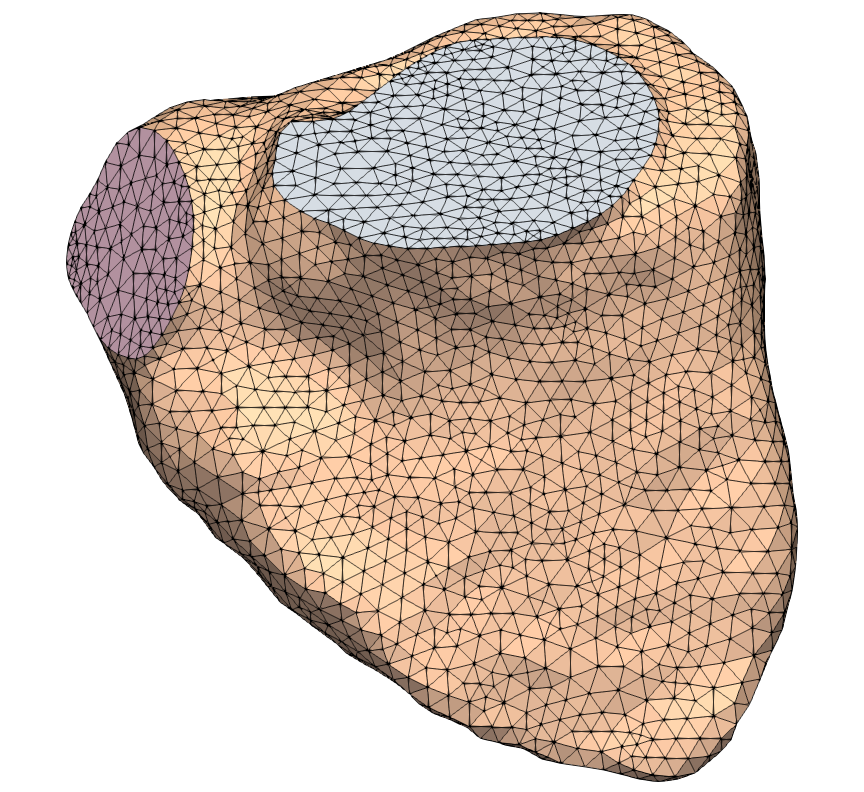}\hfill
    (b)\includegraphics[width=1.44in]{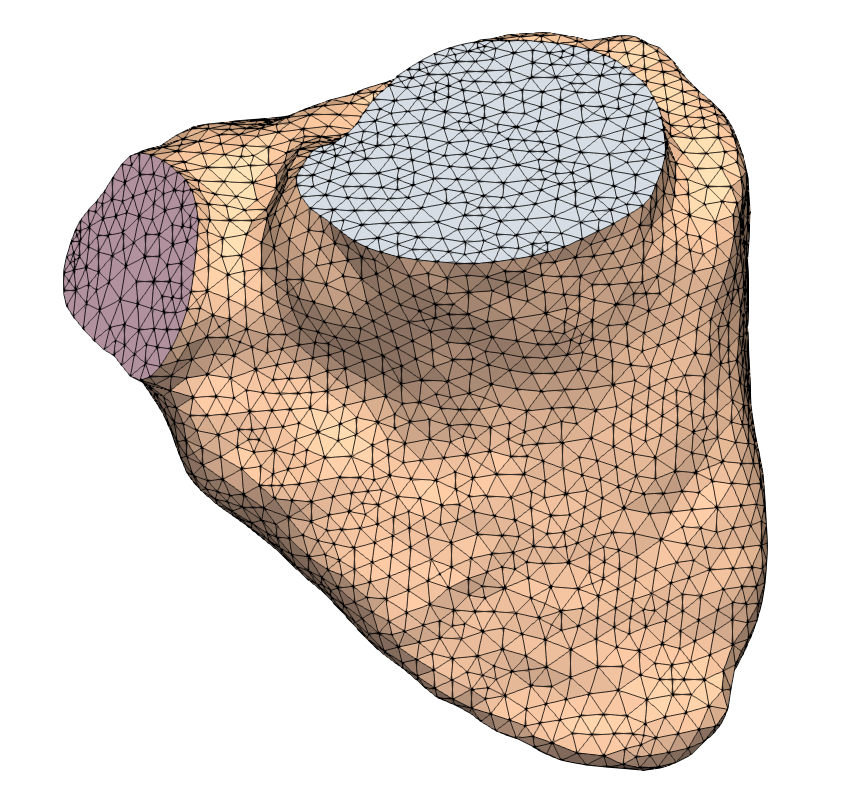}\hfill
    (c)\includegraphics[width=1.44in]{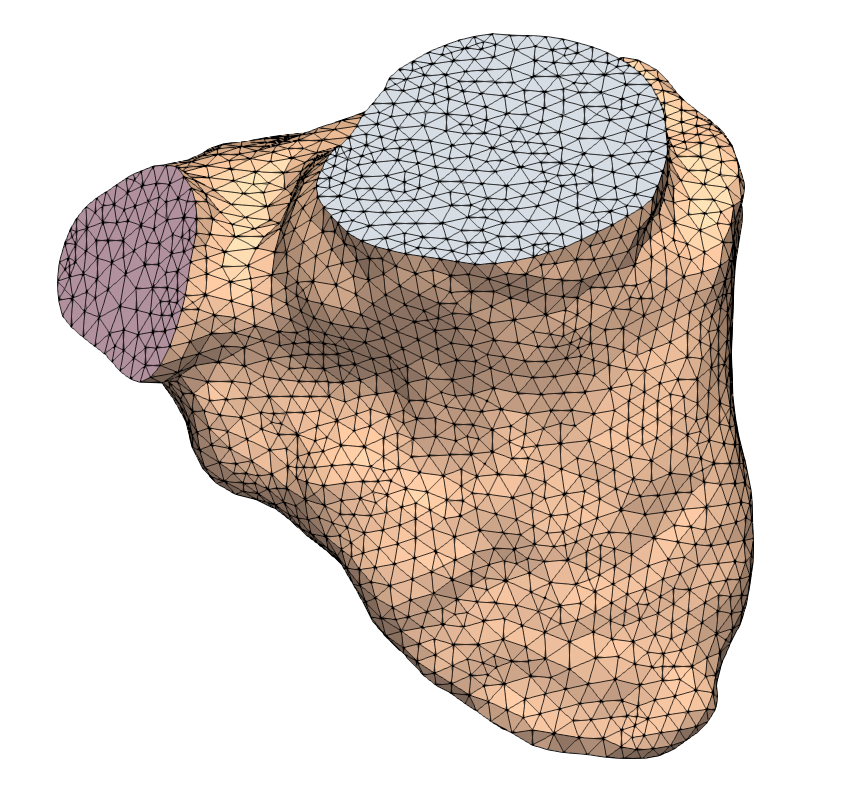}
    \caption{The right ventricle surface mesh: (a)~beginning of systole (frame 0), (b)~middle of systole (frame 23), (c)~end of systole (frame 44).}
    \label{figure:surfacemesh}
\end{figure}

At the next stage, we deform the reference mesh to sequentially adapt to all frames from 0 to 89, which is followed by the  second cycle of the mesh adaptation  to ensure a smooth periodic transition of the meshes from one cardiac cycle to the next one.  Each  step of the mesh deformation is split into two substeps. First, we move only boundary nodes {while} simultaneously propagating and smoothing the surface mesh. Each boundary node is shifted in the direction of the weighted sum of two vectors: the surface normal vector (weight 0.5) and the vector pointing at the center of surrounding nodes (weight 0.04).
This procedure is repeated until the maximum displacement drops below $\varepsilon = 0.0001$ mm, or until the maximum number of 2000 iterations is exceeded. For vertices lying on the valve planes, the displacement vectors are projected to these planes, thus ensuring the vertices stay on valve planes~(Fig.~\ref{figure:surfacemesh}). At the second step, we apply a simultaneous untangling and smoothing algorithm~\cite{Escobar20032775}; the boundary nodes are then fixed, and only the internal nodes are shifted.
The untangling stage is robust due to the presence of internal nodes in all tetrahedra.

\begin{figure}
\centering
    (a)\includegraphics[width=1.44in]{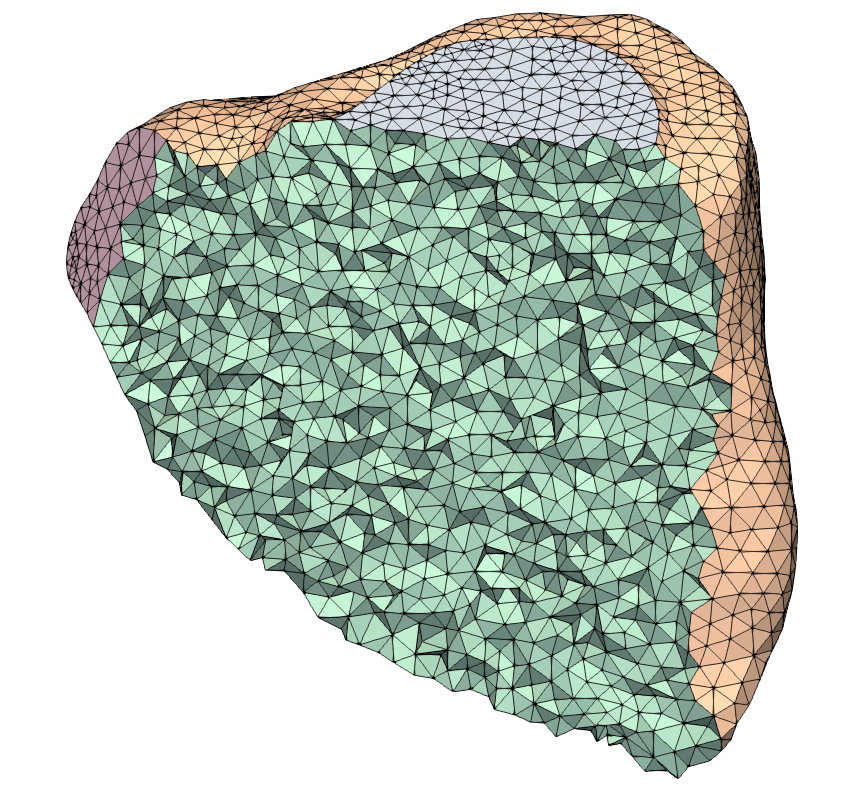}\hfill
    (b)\includegraphics[width=1.44in]{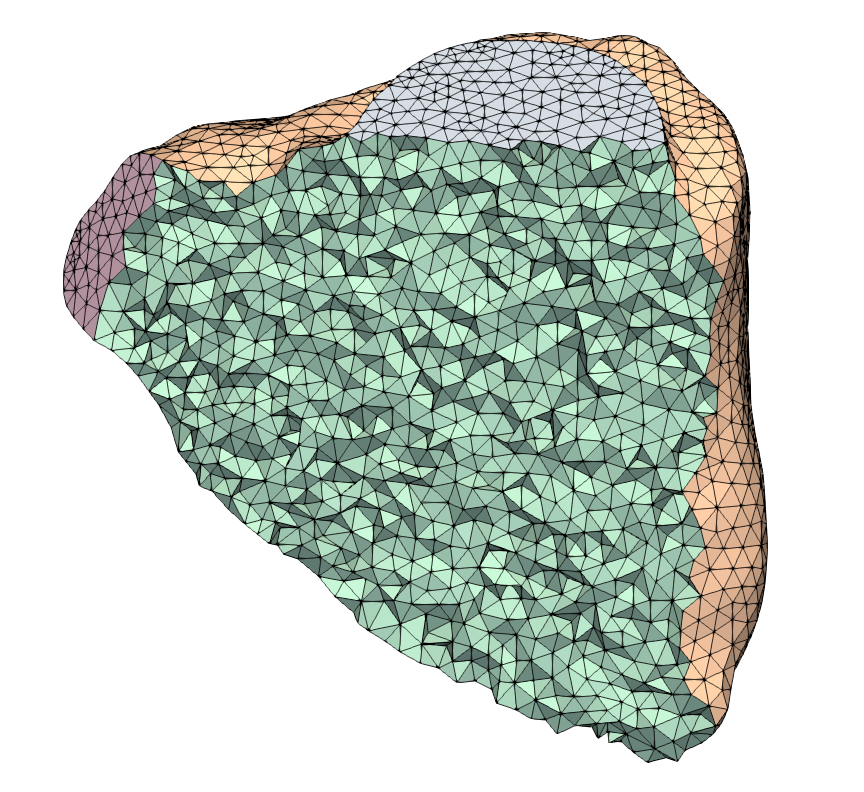}\hfill
    (c)\includegraphics[width=1.44in]{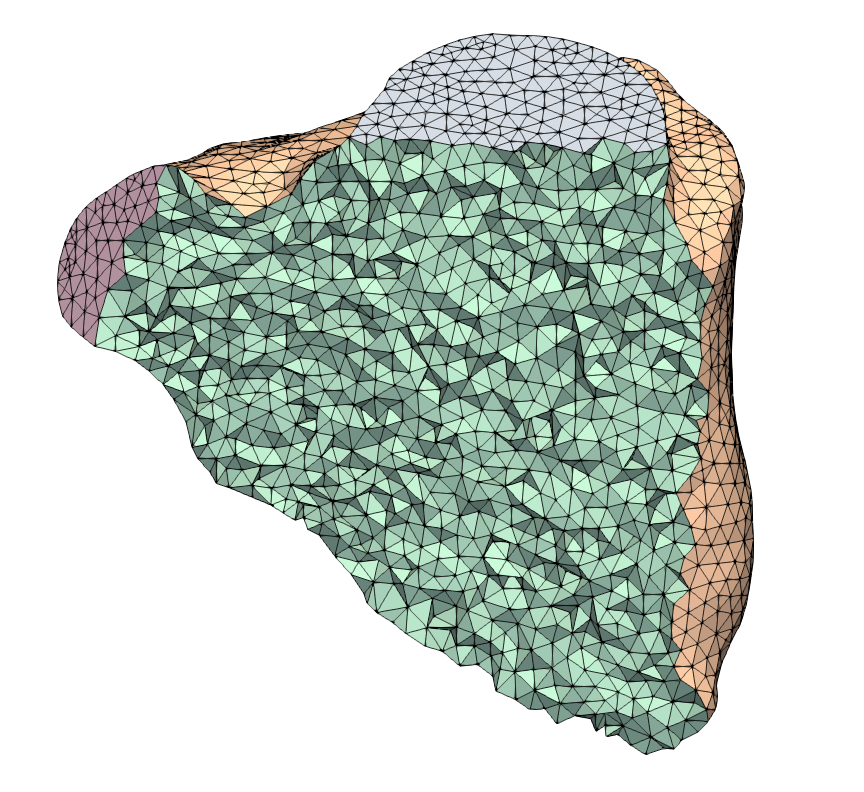}
    \caption{The right ventricle volume mesh, the cutplane passes through ventricle apex and the centers of the valves: (a)~beginning of systole (frame 0), (b)~middle of systole (frame 23), (c)~end of systole (frame 44).}
    \label{figure:volumemesh}
\end{figure}

The final result is a `periodic' series of 90 topologically invariant meshes with 13222 nodes, 86920 edges and 70533 tetrahedra for the right ventricle; see Figures~\ref{figure:surfacemesh} and~\ref{figure:volumemesh}.

\section{Numerical examples}\label{s_num_examples}

Numerical examples include convergence and stability tests for a problem with known analytical solution and
personalized simulation of the flow in the right ventricle of a post-surgery patient with TGA.
The finite element method \eqref{FE1},\eqref{SUPG},\eqref{turb_visc} is implemented within the open source Ani3D software (www.sf.net/p/ani3d).

\subsection{Convergence and stability tests}\label{s_num_test}

To set up convergence test with a synthetic solution,  we let the reference  domain $\Omega_0$ to be the axisymmetric tube defined in
cylindrical coordinates $(r, y, \phi)$ as:
$$
\Omega_0 =\{ (r, y, \phi): -4\le y \le 4, r^2\le e^{y/4+1} \}.
$$
Neumann boundary condition is set at outflow boundary $\dO_0^{\rm out} = \left\{(r, y, \phi): r\le e, y=4\right\}$, and
Dirichlet  boundary condition is set on the remaining part of the boundary $\dO_0^{\rm ns} = \dO_0 \setminus \dO_0^{\rm out}$.

The  time-dependent domain and the analytical solution $\{\bu,p\}$ to \eqref{NSE},\eqref{cont_int} are defined as follows:
$$
\Omega(t) =\left\{ (r, y, \phi): -4\le y \le 4, r^2\le e^{y/4+1}\left(1-\frac14 t\right) \right\},\quad  t\in[0,3],
$$
\begin{align*}
&u_r = -\frac{2e^{-\frac{1}{4}(y+4)}r^{3}}{(4-t)^2},\quad u_y = \frac{8}{4-t} - \frac{32 e^{-\frac{1}{4}(y+4)}r^2}{(4-t)^2},\quad u_{\phi} = 0,\\
&p = 512\nu\frac{e^{-\frac{1}{4}(y+4)}}{(4-t)^2}-8\frac{y}{(t-4)^2} + \tilde{p}(t),
\end{align*}
where $\tilde{p}(t)$ depends only on $t$. Requirement $p=0$ on $\dO^{\rm out}(t)$ guarantees the unique pressure solution.
To simplify  calculation of the corresponding body force, we replace the term $2\nu\int\limits_{\Omega(t)}\bD(\bu):\bD(\bpsi)\,dx$ in the Galerkin formulation \eqref{weak} to   $\nu\int\limits_{\Omega(t)}\nabla\bu:\nabla\bpsi\,dx$,  and in the SUPG residual of \eqref{SUPG} we use $-\nu\Delta \bu_h^k$ for the viscous term.

The right-hand side $\blf = (f_r,f_y,f_\phi)^T$ is obtained by substituting $\{\bu,p\}$ to \eqref{NSE} for the given viscosity $\nu$:
\[
f_r = \nu\frac{e^{-\frac{1}{4}(y+4)}}{(4-t)^2}\left(16r+\frac{1}{8}r^3\right)-4\frac{e^{-\frac{1}{2}(y+4)}}{(t-4)^4}r^5,\quad
f_y = 2\nu\frac{e^{-\frac{1}{4}(y+4)}}{(4-t)^2}r^2-128\frac{e^{-\frac{1}{2}(y+4)}}{(t-4)^4}r^4,\quad f_{\phi} = 0.
\]

To check the convergence of the discrete velocity and pressures, we run a series of five simulations using a sequence of unstructured quasi-uniform tetrahedral meshes with mesh sizes $h_i=2^{(1-i)/2}$, $i=1,\dots,5$. The time integration interval is  $[0, 0.2]$.
According to numerical analysis  \cite{lozovskiy2018quasi} and numerical evidence, the scheme \eqref{FE1} without convective stabilizations delivers a two-times reduction of the energy error norm if the time step $\Delta t$ decreases by two for each next mesh in this sequence. For $\bu^k:=\bu(k\Delta t),p^k:=p(k\Delta t)$, we define the error of the finite element solution $\{\bu^k_h,p^k_h\}$ as
$\{\be^k, e^k\}:= \{\bu^k-\bu^k_h,p^k-p^k_h\}$. To check the effect of the SUPG stabilization on the convergence rate, we perform  the experiment with and without the stabilization  and report the computed error norms in Table \ref{table:convergence1}.
We see that the second order asymptotic convergence  in the energy norm is observed both for the SUPG-stabilized and non-stabilized finite element schemes. The $L^2$-norm of the error demonstrates the third order convergence.
In the numerical test we set viscosity to $\nu=\frac{\mu}{\rho} = \frac{4.2\cdot 10^{-5}}{1.05\cdot 10^{-3}} = 0.04$ (all units are SI except length which is in cm) which corresponds to that of blood.
Note that the mesh size and the analytical solution are such that
the scheme \eqref{FE1} can be used without stabilization.

\begin{table}[!ht]
  \begin{center}
    \begin{tabular}{*{6}{c}}
    \hline
Mesh  size    & 1.0 & $1.0/\sqrt{2}$ & 0.5  & $0.5/\sqrt{2}$ & 0.25 \\
    \hline
Time step     & 0.04 & 0.02 & 0.01 & 0.005 & 0.0025 \\
    \hline
	    &  \multicolumn{5}{c}{no stabilization}\\
    \hline
$\underset{0\leqslant k\leqslant N}{\mbox{max}}\|\be^{k}\|$ & 0.067 & 0.038 & 0.0179 & 0.00761 & 0.00240  \\
 Error ratio  & & 1.76 & 2.1 & 2.35 & 3.17  \\
    \hline
$\sqrt{\sum_{k=1}^N\Delta t \|\nabla\be^{k}\|^2}$        & 0.203 & 0.138 & 0.0820 & 0.0466  & 0.0212   \\
 Error ratio  & & 1.47 & 1.68 & 1.76 & 2.20  \\
    \hline
	    &  \multicolumn{5}{c}{SUPG stabilization}\\
    \hline
$\underset{0\leqslant k\leqslant N}{\mbox{max}}\|\be^{k}\|$ & 0.066 & 0.037 & 0.0178 & 0.00760 & 0.00241  \\
 Error ratio  & & 1.76 & 2.1 & 2.34 & 3.15  \\
    \hline
$\sqrt{\sum_{k=1}^N\Delta t \|\nabla\be^{k}\|^2}$        & 0.200 & 0.137 & 0.0813 & 0.0464  & 0.0212   \\
 Error ratio  & & 1.46 & 1.69 & 1.75 & 2.19  \\
    \hline
    \end{tabular}
  \end{center}
\caption{Finite element errors for the given analytical solution, $\nu=0.04$. \label{table:convergence1}}
\end{table}

Now we fix the third mesh withe $h=0.5$ and perform $N=300$ time steps with $\Delta t=0.01$. This setting allows the numerical instability to develop for smaller viscosities.
In Table \ref{table:stability} we present the velocity norm
$\sqrt{\underset{0\leqslant k\leqslant N}{\mbox{max}}\frac{1}{2}\|\bu^{k}_h\|^2 + \nu\sum_{k=1}^{N}\Delta t\|\nabla\bu^{k}_h\|^2}$ for different values of viscosity. Here and further $\|\cdot\|$ denotes the $L^2$ norm.
SUPG stabilization as well as the combined SUPG/Smagorinski stabilization with $M=0.01$
produce stable solutions while the scheme \eqref{FE1} without convective stabilization blows up.

\begin{table}[!ht]
  \begin{center}
    \begin{tabular}{*{4}{c}}
    \hline
Viscosity   & $0.04$ & $0.004$& $0.0004$  \\
    \hline
no stabilization & 33.2 &  29.8 & $3.62\cdot 10^{5}$  \\
SUPG stabilization & 33.2 & 29.8  &  29.4  \\
Combined SUPG/Smagorinski stabilization & 33.2 & 29.7 & 29.3  \\
    \hline
    \end{tabular}
  \end{center}
\caption{Numeric velocity norm $\sqrt{\max\limits_{0\leq k\leq N}\frac{1}{2}\|\bu^{k}_h\|^2 + \nu\sum_{k=1}^{N}\Delta t\|\nabla\bu^{k}_h\|^2}$ over the period of $N=300$ time steps  of size $\Delta t=0.01$ each for the triangulation with mesh size $h=0.5$. \label{table:stability}}
\end{table}

In order to study the impact of the scaling given by $M$ in the definition~\eqref{turb_visc} of  the turbulent viscosity, we compute the finite element errors in the setup of the previous experiment. In Table \ref{table:errors} we show the $L^2$- and energy norm errors
of the finite element solutions stabilized by SUPG or the combined SUPG/Smagorinski method.
For relatively large viscosity, the  addition of the turbulent viscosity increases the $L^2$- and the energy error norms by 100\% and 50\%, respectively.
To the contrary, for small viscosity the partial addition of the turbulent viscosity reduces the error norms by 30\% and 40\%, respectively suggesting slightly oscillatory behavior of the pure SUPG-stabilized solution and warranting the inclusion of the weighted Smagorinski term.
\begin{table}[!ht]
  \begin{center}
    \begin{tabular}{*{4}{c}}
    \hline
Viscosity   & $0.04$ & $0.004$& $0.0004$ \\
    \hline
	    &  \multicolumn{3}{c}{SUPG stabilization}\\
    \hline
$\underset{1\leqslant k\leqslant N}{\mbox{max}}\|\be^{k}\|$ & 0.106 &   0.198  & 0.562    \\
$\sqrt{\sum_{k=1}^N\Delta t \|\nabla\be^{k}\|^2}$        & 1.00  &    2.089  & 7.00     \\
$\sqrt{\underset{0\leqslant k\leqslant N}{\mbox{max}}\frac{1}{2}\|\be^{k}\|^2 + \nu\sum_{k=1}^{N}\Delta t\|\nabla\be^{k}\|^2}$ & 0.214 & 0.192 & 0.422 \\
    \hline
	    &  \multicolumn{3}{c}{Combined SUPG/Smagorinski stabilization}\\
    \hline
$\underset{0\leqslant k\leqslant N}{\mbox{max}}\|\be^{k}\|$ & 0.198 & 0.328 & 0.396    \\
$\sqrt{\sum_{k=1}^N\Delta t \|\nabla\be^{k}\|^2}$        & 1.47  & 3.08  & 4.24     \\
$\sqrt{\underset{0\leqslant k\leqslant N}{\mbox{max}}\frac{1}{2}\|\be^{k}\|^2 + \nu\sum_{k=1}^{N}\Delta t\|\nabla\be^{k}\|^2}$ & 0.326 & 0.303 & 0.293 \\
    \hline
    \end{tabular}
  \end{center}
\caption{Finite element errors for the given analytical solution after $N=300$ time steps with $\Delta t=0.01$ and $h=0.5$. \label{table:errors}}
\end{table}

\subsection{Blood flow in the right ventricle}\label{s_tga}
We now apply the finite element methods with Smagorinski, SUPG  and the combined stabilization to simulate blood flow in  the right ventricle. For this purpose we use a sequence of topologically equivalent meshes described in  section~\ref{sec_seg}.
The duration of the cardiac cycle is $0.87s$. Since we have 90 time frames, the time step $\Delta t$ in \eqref{FE1} was set to be equal to $0.87/90s$. Boundary conditions were set according to \eqref{cont_int}--\eqref{bc3}, with $\bxi_t$ computed on the boundary by the first order finite-difference stencil directly from the displacement at the nodes and edges.
Numerical simulations start from the flow at rest as the initial condition and run over two cardiac cycles. All results below are shown for the second cycle.

\begin{figure}
\centering
    \begin{overpic}[width=.25\textwidth,grid=false]{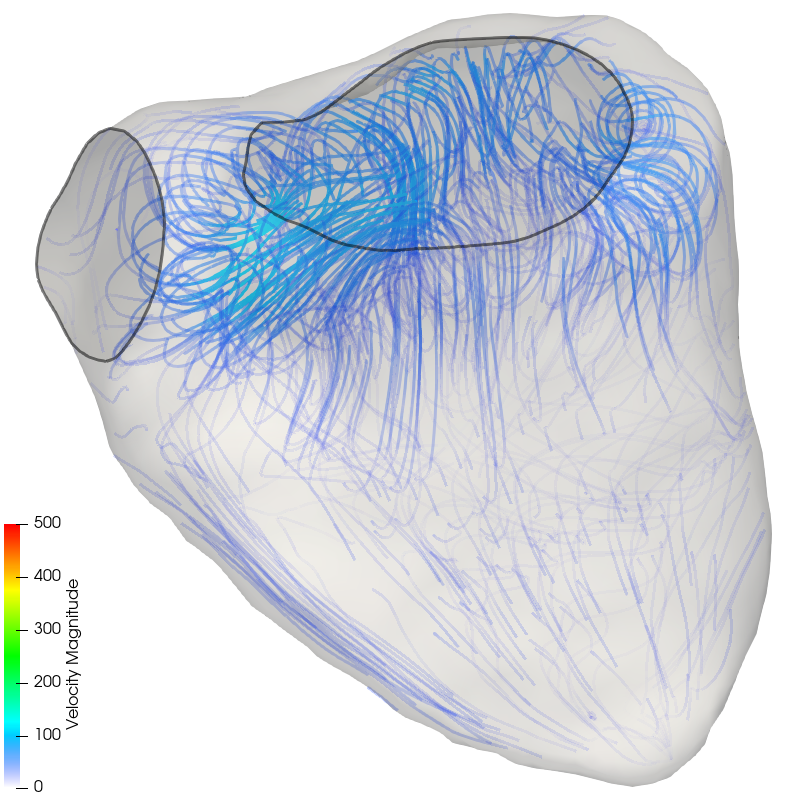}
    \put(30,102){\small{Smagorinski}}
    \end{overpic}\hfill
    \begin{overpic}[width=.25\textwidth,grid=false]{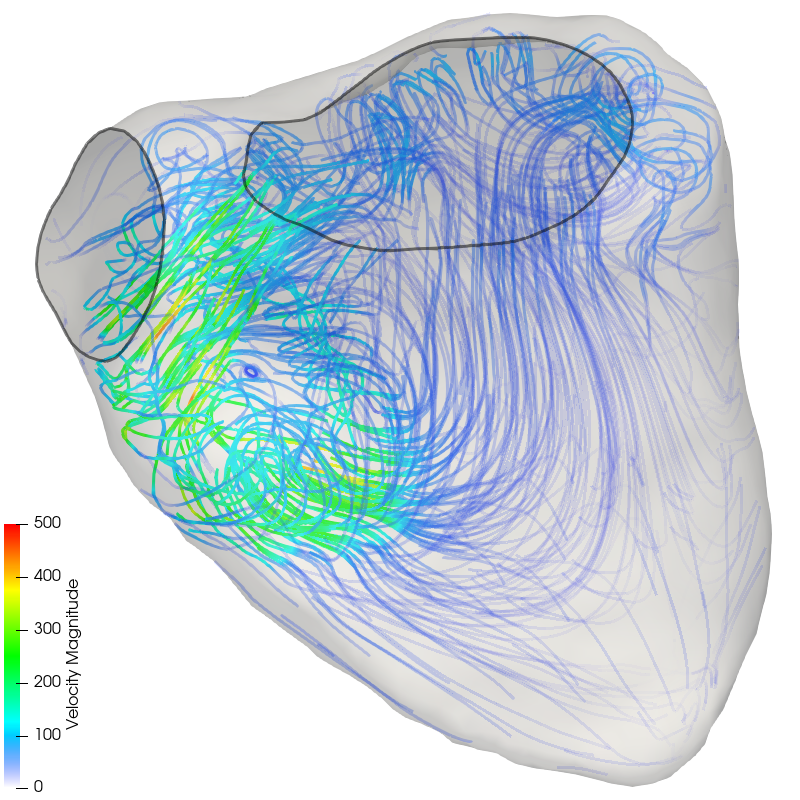}
    \put(0,102){\small{0.01*Smagorinski+SUPG}}
    \end{overpic}\hfill
    \begin{overpic}[width=.25\textwidth,grid=false]{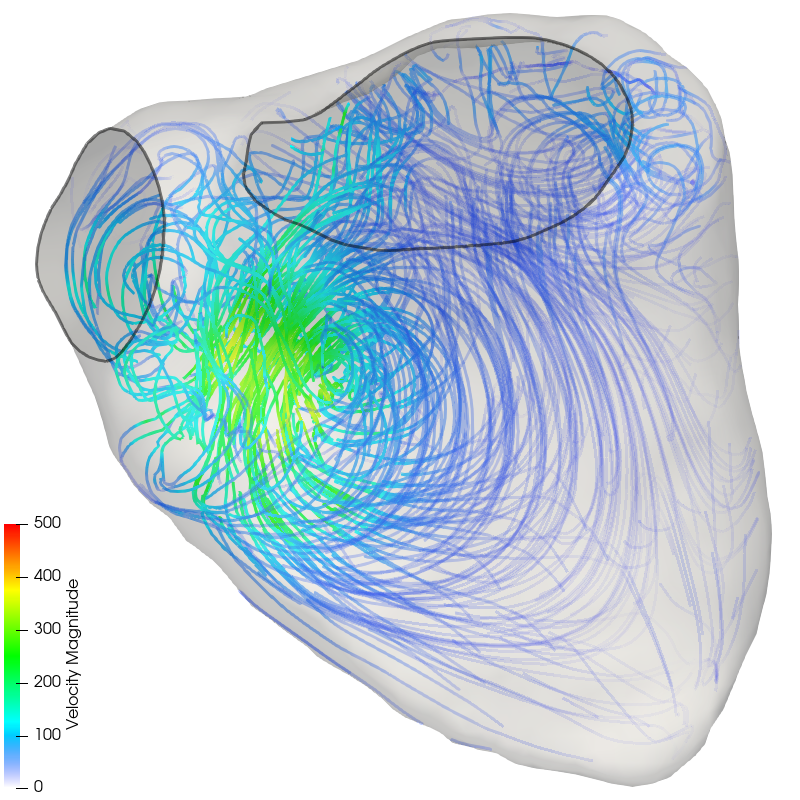}
    \put(37,102){\small{SUPG}}
    \end{overpic}\\[5ex]
    \begin{overpic}[width=.25\textwidth,grid=false]{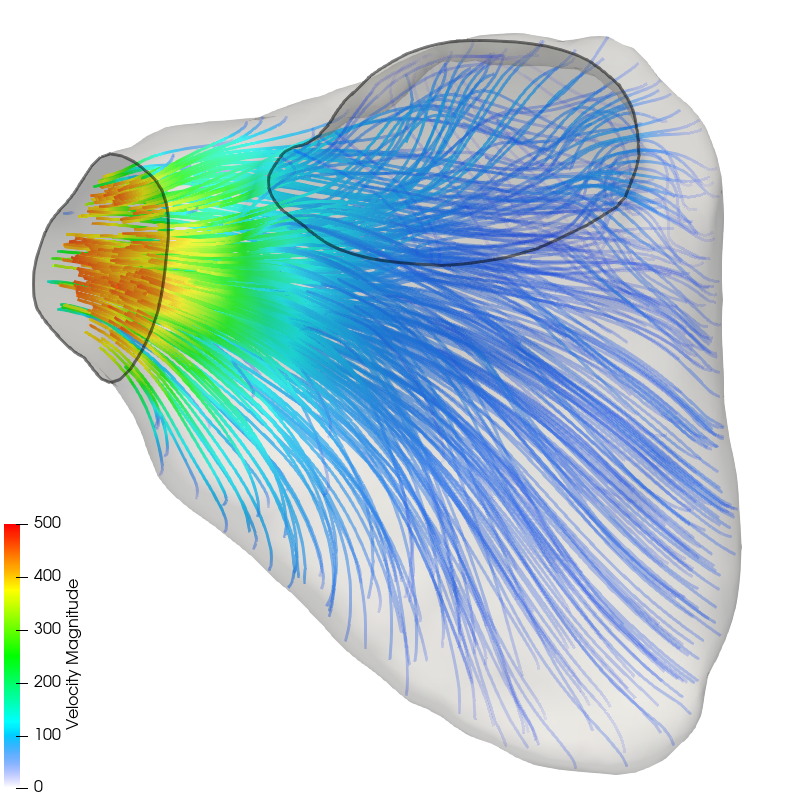}
    \put(37,102){\small{Smagorinski}}
    \end{overpic}\hfill
    \begin{overpic}[width=.25\textwidth,grid=false]{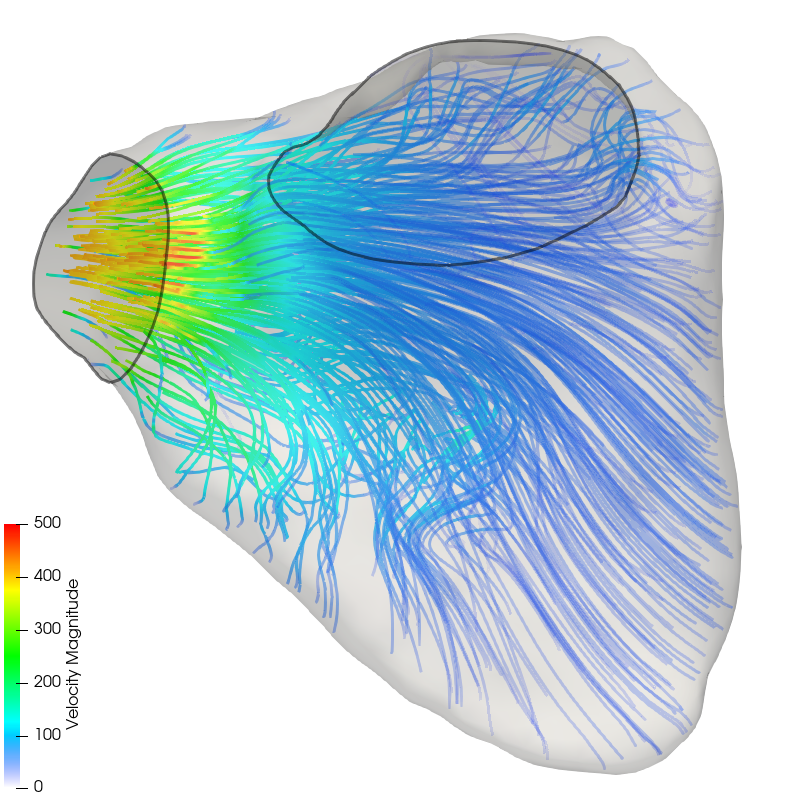}
    \put(0,102){\small{0.01*Smagorinski+SUPG}}
    \end{overpic}\hfill
    \begin{overpic}[width=.25\textwidth,grid=false]{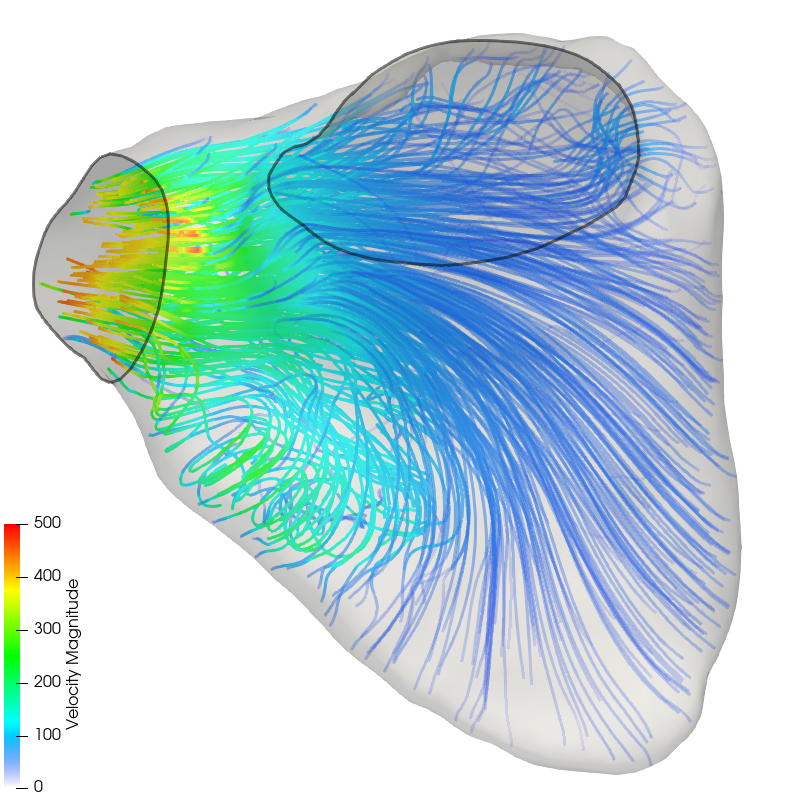}
    \put(30,102){\small{SUPG}}
    \end{overpic}
    \caption{Visualization of streamlines. The upper row is for the beginning of systole. The bottom row is the middle of systole.}
    \label{figure:stream}
\end{figure}

From the numerical results for the problem with the synthetic solution,
we see that both  SUPG and the combination of  weighted Smagorinski and SUPG stabilizations deliver convergent solutions with the pure SUPG
being closer to the borderline of numerically stable simulations.
Pure Smagorinski stabilization is a well-known turbulent LES model that, however, may over-dissipate.
We are now interested in how these properties affect the patient specific  CFD visualization of blood flow in the right ventricle.

\begin{figure}
\centering
    \begin{overpic}[width=.25\textwidth,grid=false]{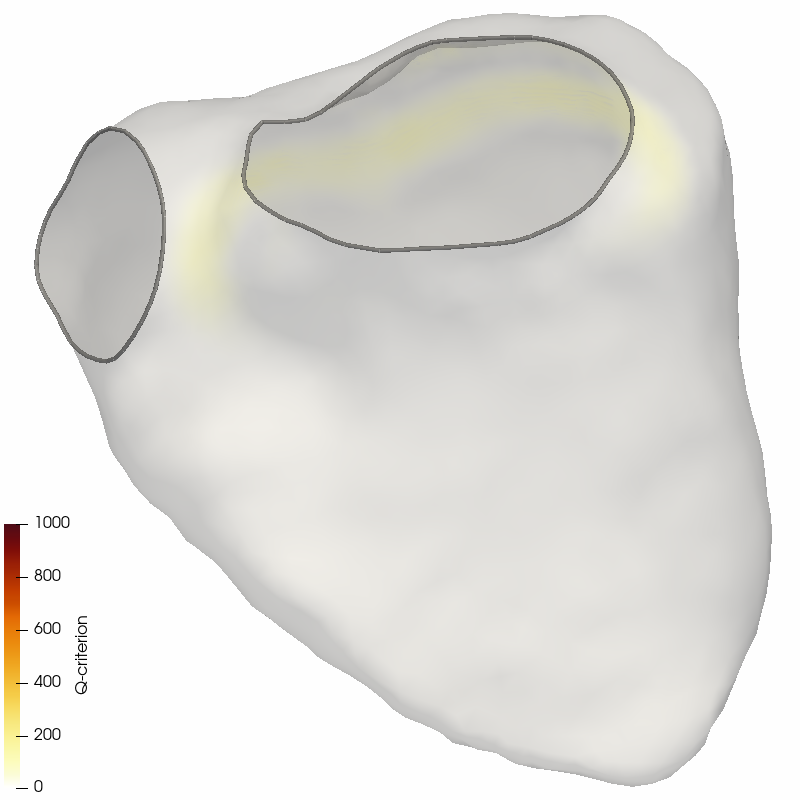}
    \put(30,102){\small{Smagorinski}}
    \end{overpic}\hfill
    \begin{overpic}[width=.25\textwidth,grid=false]{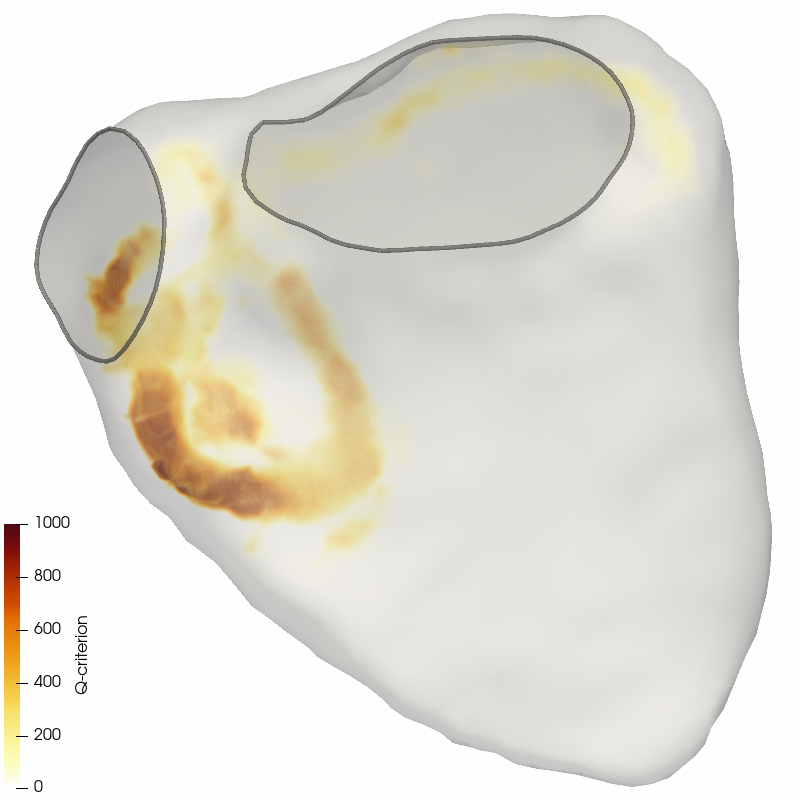}
    \put(0,102){\small{0.01*Smagorinski+SUPG}}
    \end{overpic}\hfill
    \begin{overpic}[width=.25\textwidth,grid=false]{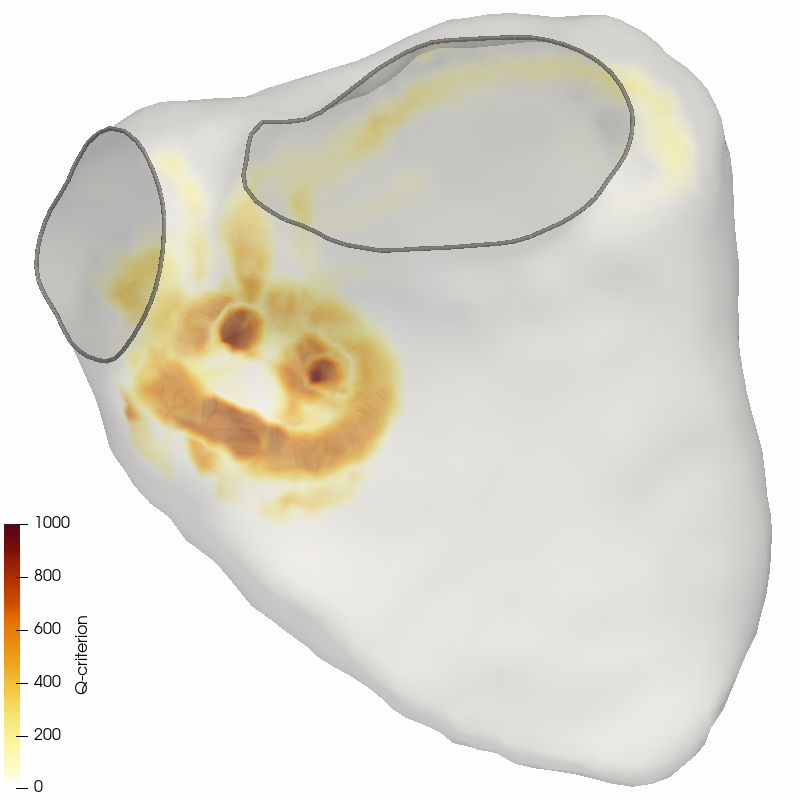}
    \put(37,102){\small{SUPG}}
    \end{overpic}\\[5ex]
    \begin{overpic}[width=.25\textwidth,grid=false]{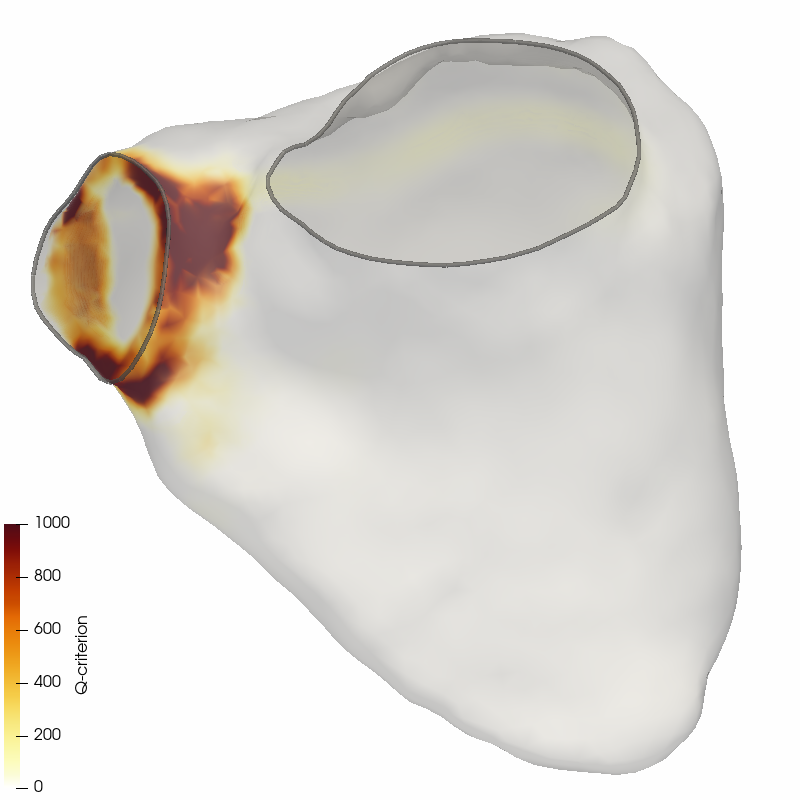}
    \put(37,102){\small{Smagorinski}}
    \end{overpic}\hfill
    \begin{overpic}[width=.25\textwidth,grid=false]{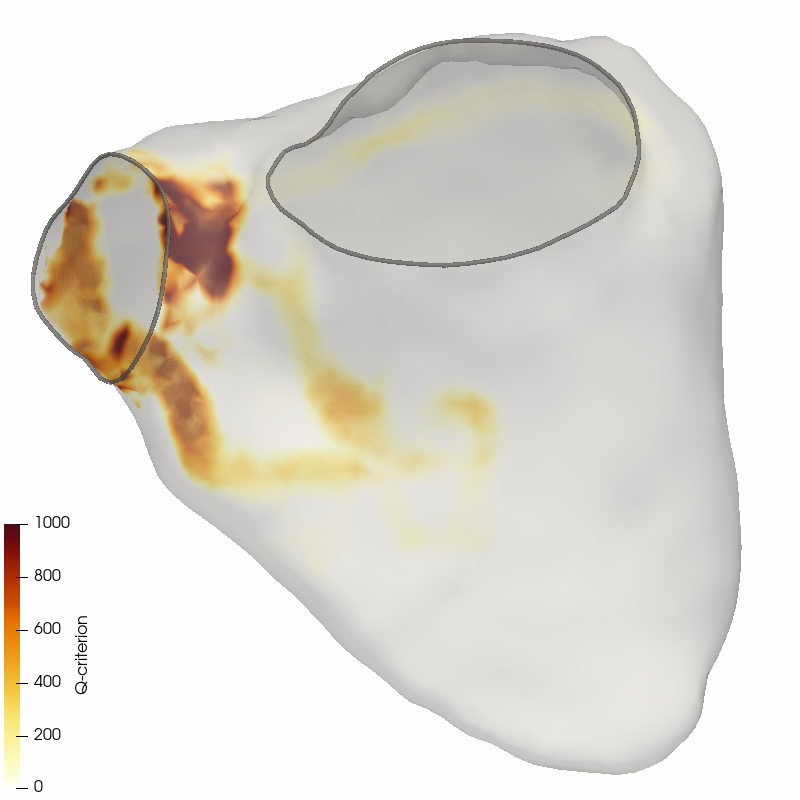}
    \put(0,102){\small{0.01*Smagorinski+SUPG}}
    \end{overpic}\hfill
    \begin{overpic}[width=.25\textwidth,grid=false]{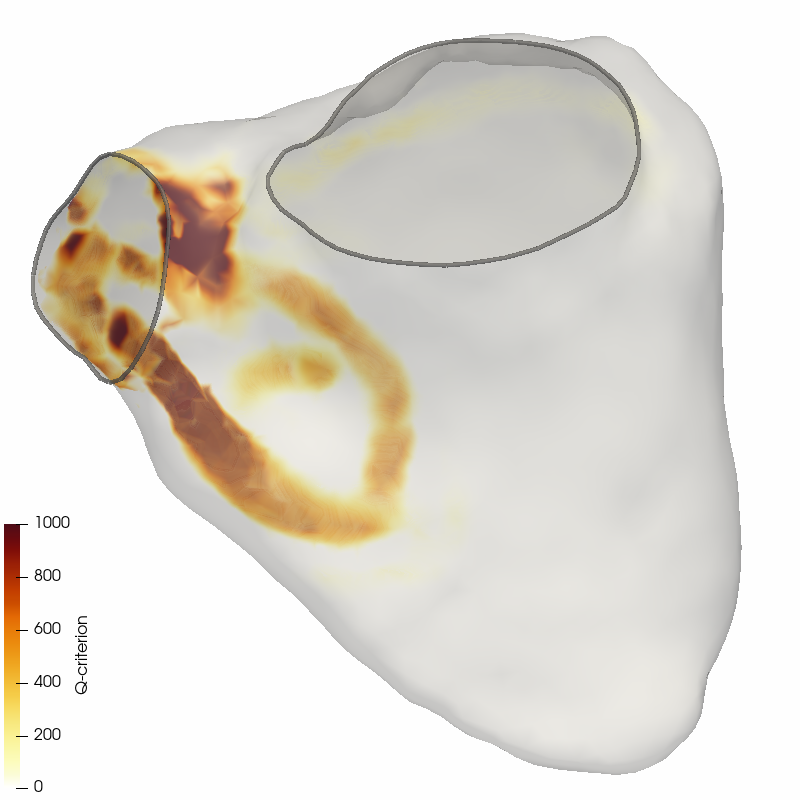}
    \put(30,102){\small{SUPG}}
    \end{overpic}
    \caption{Visualization of vortex regions according to the Q-criterion. The upper row is for the beginning of systole. The bottom row is the middle of systole.}
    \label{figure:q-crit}
\end{figure}

In Figure~\ref{figure:stream} we show streamlines that are computed as particle tracers in blood flow at a given time.
The flow is visualized at the  beginning and middle of the systole. 
We see that among three models the pure SUPG and the weighted Smagorinski combined with SUPG stabilizations 
capture the large recirculation zone in front of the pulmonary valve.
This observation is confirmed by the picture of the $Q$-criterion levels in Figure~\ref{figure:q-crit}:
the zones where $Q=\frac12(\text{tr}^2(\nabla\bu) - \text{tr}([\nabla\bu]^2))$ 
is positive are identified as  vortexes in  incompressible flow~\cite{dubief2000coherent}.
Again, pure SUPG model suggests most distinct vortical regions and using standard Smagorinski model suppresses vorticity.
The combination of SUPG and weighted Smagorinski model is doing a reasonably good job in identifying the vortical region.
All these observations confirm that standard Smagorinski stabilization is over-diffusive for numerical simulations of the flow in the heart.

\begin{figure}
\centering
    \begin{overpic}[width=.25\textwidth,grid=false]{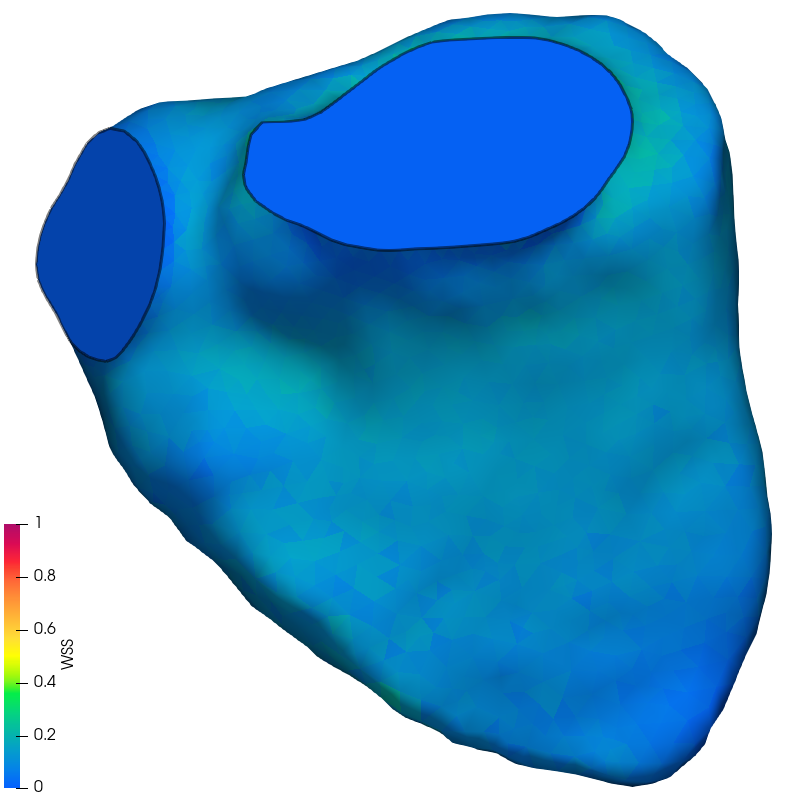}
    \put(30,102){\small{Smagorinski}}
    \end{overpic}\hfill
    \begin{overpic}[width=.25\textwidth,grid=false]{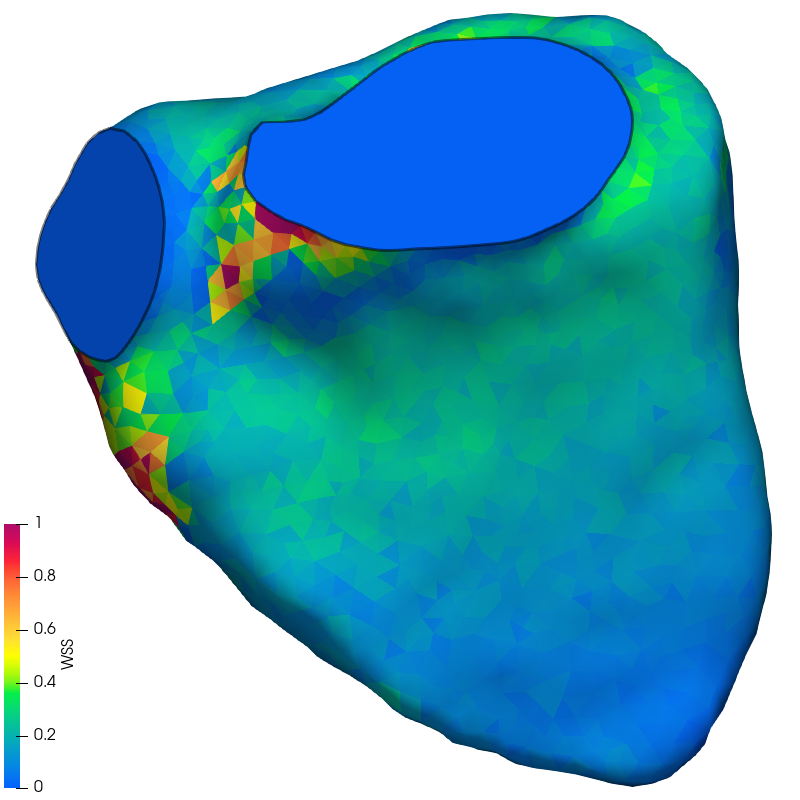}
    \put(0,102){\small{0.01*Smagorinski+SUPG}}
    \end{overpic}\hfill
    \begin{overpic}[width=.25\textwidth,grid=false]{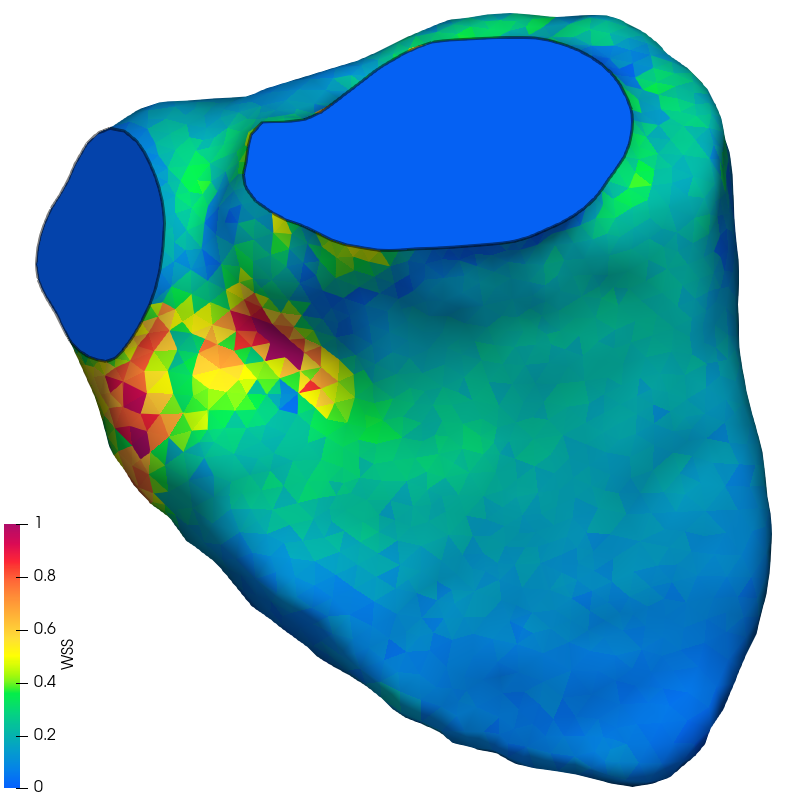}
    \put(37,102){\small{SUPG}}
    \end{overpic}\\[5ex]
    \begin{overpic}[width=.25\textwidth,grid=false]{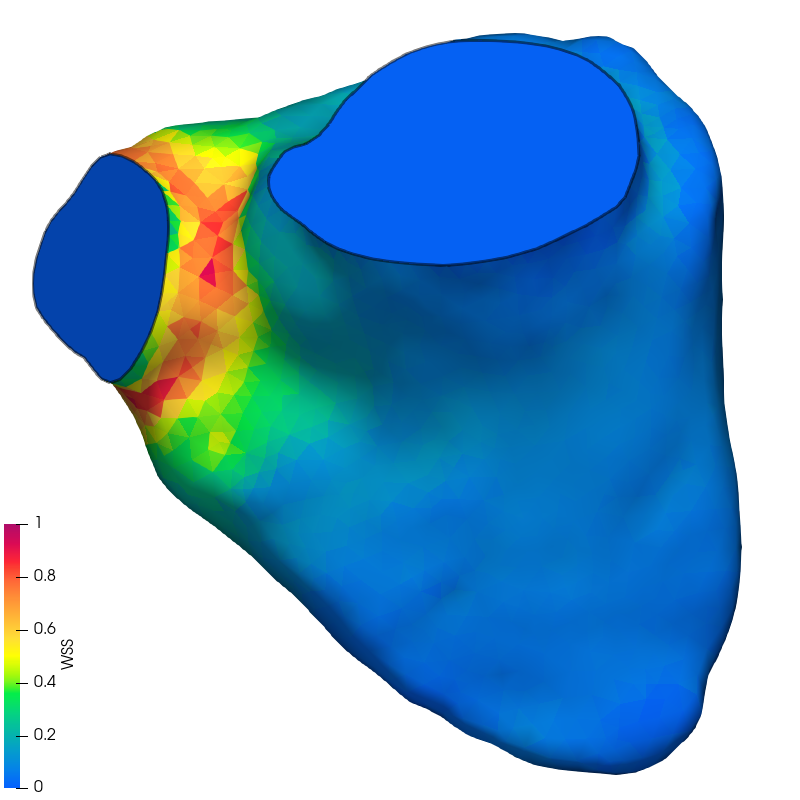}
    \put(37,102){\small{Smagorinski}}
    \end{overpic}\hfill
    \begin{overpic}[width=.25\textwidth,grid=false]{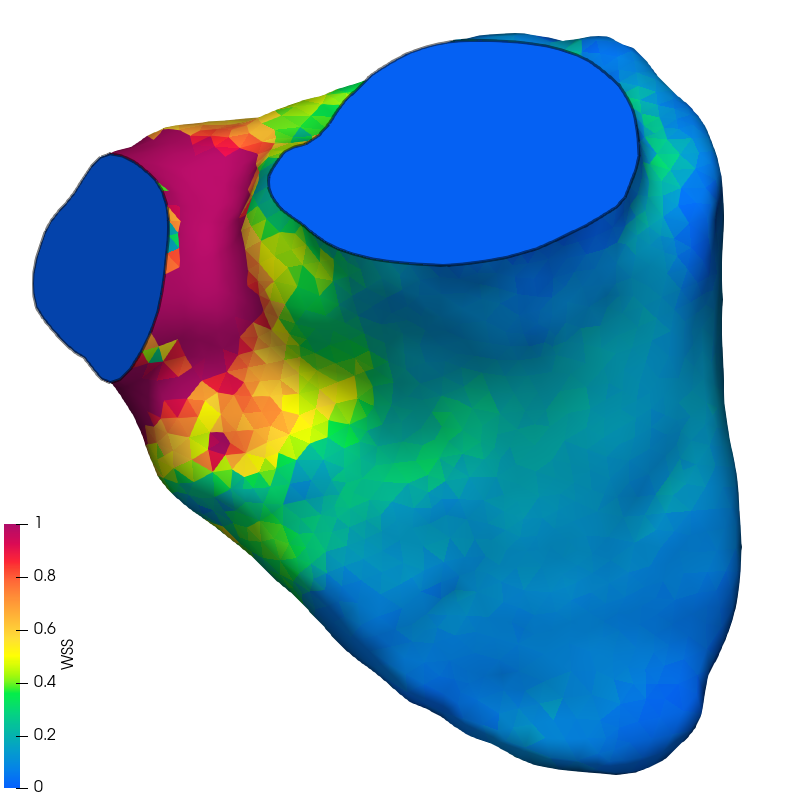}
    \put(0,102){\small{0.01*Smagorinski+SUPG}}
    \end{overpic}\hfill
    \begin{overpic}[width=.25\textwidth,grid=false]{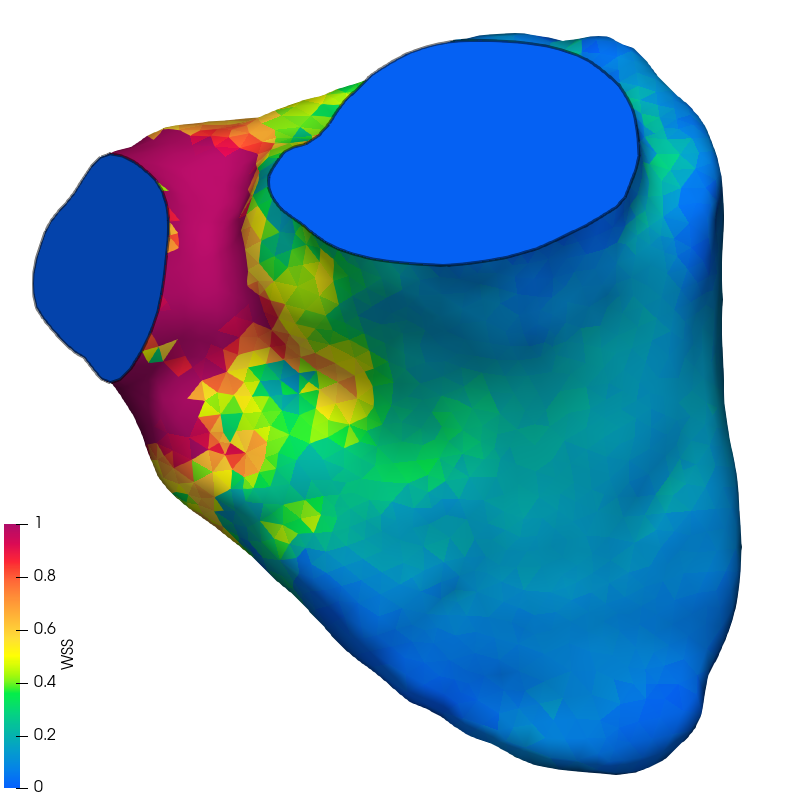}
    \put(30,102){\small{SUPG}}
    \end{overpic}
    \caption{Visualization of the wall shear stress. The upper row is for the beginning of systole. The bottom row is the middle of systole.}
    \label{figure:wss}
\end{figure}

Finally, we calculate the wall shear stress (WSS) predicted by the three models and show them in Figure~\ref{figure:wss}.
The WSS is likely underestimated by the Smagorinski model, which is a consequence of its overall poor accuracy. Less diffusive methods predict larger wall shear stress.

We conclude that both SUPG and the combination of SUPG and weighted Smagorinski stabilizations of the finite element method can be used
in a predictive CFD visualization of cardiac flows with SUPG being the least diffusive, while SUPG/weighted Smagorinski  adds extra stability.\\

Acknowledgement. The work was supported by the Russian Science Foundation grant 19-71-10094 
(A.Danilov, A.Lozovskiy, V.Salamatova, implementation and numerical experiments), 
the Moscow Center for Fundamental and Applied Mathematics, agreement  No. 075-15-2019-1624 (Yu.Vassilevski, problem setting and discrete models).
Maxim Olshanskii (problem setting and discrete models) was partially supported by NSF through DMS-1953535 and DMS-2011444.


\bibliographystyle{elsarticle-num}
\bibliography{mybib}

\end{document}